\documentclass{aa}  

\usepackage{natbib}
\usepackage{graphicx}
\usepackage{txfonts}
\usepackage[pdftex]{hyperref}
\hypersetup{urlcolor=black, citecolor=blue, linkcolor=blue, colorlinks=true}  
\usepackage{amsmath}
\usepackage{amssymb}
\usepackage{bm}
\usepackage{balance}

\begin{document}
    \titlerunning{A search for X-ray absorbed sources in the 3XMM catalogue}
    \title{A search for X-ray absorbed sources in the 3XMM catalogue, 
           using photometric redshifts and Bayesian spectral fits}

    \author{A. Ruiz
            \inst{1}        
        \and
            I. Georgantopoulos
            \inst{1}
        \and
            A. Corral
            \inst{2}
    }

    \institute{  
        Institute for Astronomy, Astrophysics, Space Applications, and Remote Sensing (IAASARS), 
	    National Observatory of Athens,
	    15236 Penteli, Greece \\
        \email{ruizca@noa.gr}	      
     \and   
        Instituto de Física de Cantabria (UC-CSIC),
        Av. de los Castros s/n, 39005, Santander, Spain
    }

   \date{Received September 14, 2020; accepted November 11, 2020}

  \abstract{
  Since its launch in 1999, the XMM-\textit{Newton} mission has compiled the 
  largest catalogue of serendipitous X-ray sources, with the 3XMM being the third
  version of this catalogue. This is because of the combination of a large 
  effective area (5000 $\rm cm^2$ at 1 keV) and a wide field of view (30 arcmin).
  The 3XMM-DR6 catalogue contains about 470\,000 
  unique X-ray sources over an area of 982 $\rm deg^2$.  A significant fraction 
  of these (100\,178 sources) have reliable optical, near/mid-IR counterparts 
  in the SDSS, PANSTARRS, VIDEO, UKIDSS and WISE surveys. In a previous paper 
  we have presented photometric redshifts for these sources using the TPZ machine 
  learning algorithm. About one fourth of these (22\,677) have adequate photon 
  statistics so that a reliable X-ray spectrum can be extracted. Obviously, 
  owing to both the X-ray counts selection and the optical counterpart constraint, 
  the sample above is biased towards the bright sources. Here, we present XMMFITCAT-Z: 
  a spectral fit catalogue for these sources using the Bayesian X-ray Analysis (BXA) 
  technique. As a science demonstration of the potential of the present catalogue,
  we comment on the optical and mid-IR colours of the 765 X-ray absorbed sources 
  with $N_\mathrm{H} > 10^{22}\,\mathrm{cm}^{-2}$. We show that a considerable 
  fraction of X-ray selected AGN would not be classified as AGN following the 
  mid-IR W1--W2 vs. W2 selection criterion. These are AGN with lower luminosities,
  where the contribution of the host galaxy to the MIR emission is non-negligible.
  Only one third of obscured AGN in X-rays present red colours or r--W2 > 6. 
  Then it appears that the r--W2 criterion, often used in the literature for the 
  selection of obscured AGN, produces very different X-ray absorbed AGN samples
  compared to the standard X-ray selection criteria. 
  }

   \keywords{Catalogs -- X-rays: galaxies -- Galaxies: active}

   \maketitle
%

\section{Introduction}
\label{sec:intro}
The XMM-\textit{Newton} mission \citep{Jansen01} launched in 1999 is the 
European Space Agency's (ESA) second cornerstone mission. It carries onboard 
three telescopes with the largest effective area in a X-ray telescope so far 
(combined 5000 cm$^2$ at 1 keV). These telescopes focus the light on the 
European Photon Imaging Camera (EPIC) CCD cameras. Owing to its large 
field-of-view ($\sim30$ arcmin diameter), it can detect a large number of 
serendipitous sources in each observation. The Survey Science Centre (SSC), a 
consortium of ten European Institutes, was formed and is responsible for 
producing the catalogues of the serendipitous sources detected. The third 
version of this catalogue (3XMM) has been published and is described in detail 
in \citet{3xmm}. The 3XMM covers about 1000 square degrees and contains about
half a million unique sources. This is the largest catalogue of X-ray sources 
ever produced. Their median flux of the sources is 
$\sim 2.4\times 10^{-14}\,\mathrm{erg\, cm^ {-2}\, s^{-1}}$ in the total 
0.2--12 keV band.

The large sky area of the 3XMM offers the opportunity of identify rare objects 
that are harder to find in deeper surveys with much smaller sky areas, like 
high luminosity active galactic nuclei (AGN). Moreover, the sensitivity and 
energy range of the 3XMM allows the detection of X-ray absorbed AGN, which are 
elusive in shallower and/or softer surveys. In addition, the SSC provides 
spectral products for a large number of 3XMM sources, allowing a systematic 
study of the spectral properties of the 3XMM.

The goal of this work was the construction of a clean, reliable sample of 
X-ray absorbed AGN, after a robust, systematic analysis of their X-ray spectra 
included in the 3XMM. Given the size of the available data (more than 170\,000 
spectra in the sixth version of the catalogue, 3XMM-DR6), the use of automated 
spectral modelling techniques is mandatory. The XMM-\textit{Newton} spectral-FIT 
CATalogue \citep[XMMFITCAT;][]{xmmfitcat} was a previous effort in this 
direction. This catalogue offers spectral fits for a large fraction of 3XMM 
sources having spectral data and it has been used to identify X-ray absorbed 
sources \citep{Corral14}. However, due to the lack of redshift information for 
the majority of 3XMM sources, it was not possible to provide physical, reliable 
estimations for the X-ray absorption (i.e. Hydrogen column density) or other 
spectral parameters and derived quantities (e.g. temperature of plasmas or 
luminosities).

Obtaining spectroscopic redshifts for the bulk of the 3XMM catalogue would be 
extremely costly in observational terms. Therefore, using photometric redshifts 
is the only feasible technique to obtain distance information for such a large 
number of sources. The XMM-\textit{Newton} Photo-Z CATalogue (XMMPZCAT) offers 
photometric redshifts for a large fraction of 3XMM sources: In the framework of
the ESA Prodex project we derived photometric redshifts for 100\,178 X-ray 
sources, about 50\% of the total number of 3XMM sources (205\,380) in the 
XMM-\textit{Newton} fields selected to build XMMPZCAT (4208 out of 9159). The 
photometric redshifts were derived using a machine learning algorithm, Trees 
for Photo-Z \citep[TPZ;][]{tpz}. The optical photometry used for the derivation 
of photometric redshifts included the SDSS survey and the PANSTARS survey, with 
ancillary data in near- (2MASS, UKIDSS, VISTA) and mid-IR (All-WISE survey) bands.
See \citet{xmmpzcat} for a complete description of this catalogue.

In this work we combine the results of XMMPZCAT with the automated spectral 
fitting approach of XMMFITCAT in order to estimate X-ray models and the 
corresponding spectral properties for 30\,816 3XMM detections. These results
are collected in the XMM-Newton spectral-fit z catalogue (XMMFITCAT-Z).

As a demonstration of the scientific potential of XMMFITCAT-Z, we searched
for sources showing spectral features of X-ray absorbed AGN. We found 1421 
spectra with signs of X-ray absorption ($\sim 5\%$ of the total catalogue),
corresponding to 1037 unique sources. After further examination of their
X-ray properties, we selected in our final sample 977 detections,
corresponding to 765 unique sources, showing highly reliable absorption
features and redshift estimations.

The structure of the paper is as follows: in Sect.~\ref{sec:fitting} we 
describe the XMMFITCAT-Z, including the data and techniques to build the 
catalogue. Section~\ref{sec:xrayabs} explains our methodology to identify 
X-ray absorption and the final atlas of  X-ray absorbed AGN. In 
Sect.~\ref{sec:discuss} we discuss our results, while 
Sect.~\ref{sec:conclusion} summarizes the main conclusions of this work.

\section{XMMFITCAT-Z: A catalogue of automatic X-ray spectral fitting}
\label{sec:fitting}
In this work we present the XMM-\textit{Newton} spectral-fit z 
catalogue\footnote{\url{http://xraygroup.astro.noa.gr/Webpage-prodex/xmmfitcatz_access.html}} 
(XMMFITCAT-Z): using the pipeline spectral products from the 3XMM catalogue
(see Sect.~\ref{sec:xmm}) and the redshifts we derived in XMMPZCAT (see 
Sect.~\ref{sec:photz}), we built a catalogue of X-ray spectral fits, similar 
to XMMFITCAT \citep{xmmfitcat} but now including the distance information in an 
automated spectral fitting process. This allows the estimation of important 
physical parameters from the X-ray spectra like rest-frame column densities, 
temperature of emitting plasma, intrinsic fluxes and luminosities.

XMMFITCAT-Z contains spectral-fitting results for 30\,816 source detections,
corresponding to 22\,677 unique sources. The version of the catalogue used to 
construct the current catalogue is 3XMM-DR6. Not all detections with extracted 
spectra are included in this database. To ensure the reliability of the 
spectral-fitting results, a constrain has been imposed to the number of counts 
collected in the X-ray spectra (Sect.~\ref{sec:xmm}) and as result, some 
detections are rejected in the automated spectral-fit pipeline. Moreover, only detections with redshifts in XMMPZCAT have been included (Sect.~\ref{sec:photz}). 
We offer two main products to the final user: A  single FITS table containing 
the most useful data from XMMFITCAT-Z, and a relational database with all 
information. 

The main results of the XMMFITCAT-Z database are summarized in a table in FITS format.
It contains one row for each detection included in the database, and 157 columns
containing information about the source detection and the spectral-fitting results.
The first 14 columns contain information about the source and observation, including 
redshift information, whereas the remaining 143 columns contain, for each model 
applied, model spectral-fit flags, parameter values and errors, fluxes, 
luminosities and five columns to describe the goodness of the fit. See 
Appendix~\ref{catalogue} for a full description of all the columns included in 
the table.

We also offer the full XMMFITCAT-Z database as a file in SQLite format that can 
be downloaded by the final user. The SQLite version contains significantly more 
information than the FITS table and allows performing advanced queries using SQL 
language. It is composed of 11 tables that store all the information about the 
detections, the spectra used in the automatic fitting and the results: fit 
statistics, best-fit parameters, fluxes and luminosities. A complete description 
of all tables and parameters is available in the XMMFITCAT-Z 
web-site.\footnote{\url{http://xraygroup.astro.noa.gr/Webpage-prodex/xmmfitcatz_sql_schema}}

\begin{figure}[t]
  \centering
  \resizebox{\hsize}{!}{\includegraphics{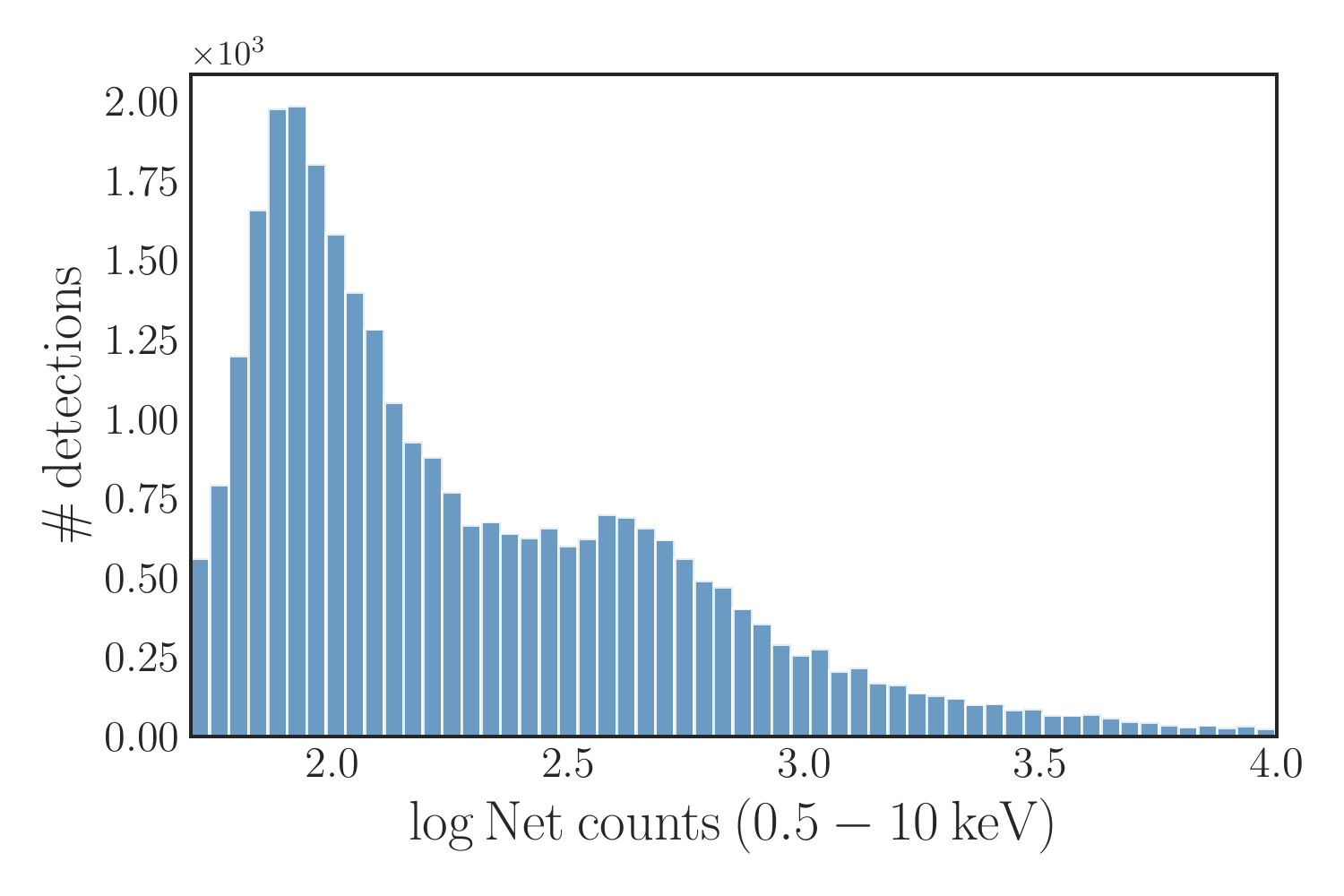}}
  \caption{Distribution of net counts (background subtracted spectral counts)
           per observation used in the automated spectral fits. Observations
           with more than 10,000 counts are not included in this plot.}
\label{fig:xfcz_counthist}
\end{figure}

\subsection{Data}
\label{sec:data}

\subsubsection{XMM-\textit{Newton} data}
\label{sec:xmm}
3XMM-DR6 is the third generation catalogue of serendipitous X-ray sources 
from the ESA's XMM-\textit{Newton} observatory, and has been created by the 
XMM-\textit{Newton} SSC on behalf of ESA. The 3XMM-DR6 catalogue contains 
X-ray photometric information for 678\,680 X-ray detections. 

Each XMM-\textit{Newton} observation considered for the construction of the 3XMM goes 
through the SSC source detection pipeline independently, obtaining a list of detections
for each observation. Since some of these observations correspond to overlapping sky
areas, a fraction of these detections can be associated to the same physical object
(i.e. unique source). The 3XMM catalogue groups as a single entry these detections
that are assumed to be the same physical source. When taking this into account, the
678\,680 detections included in the 3XMM-DR6 are reduced to 468\,440 unique sources. 

As part of the catalogue production, spectra and time series are also
extracted in the case of detections with more than 100 counts collected in the 
EPIC camera. This corresponds to 149\,998 detections, and 102\,157 unique sources.
The spectral data from the 3XMM-DR6 catalogue were screened before applying our 
automated spectral-fitting pipeline so that spectral fits are only performed if 
the data fulfil the following criteria:
\begin{itemize}
    \item Only spectra corresponding to a single instrument and observation 
    with more than 50 net counts (i.e. background subtracted) in the 0.5--10 keV 
    band are used in the spectral fits. This means that some detections with 
    more than 100 EPIC counts in the total band, but less than 50 counts in 
    each different EPIC instrument (EPIC-pn, EPIC-MOS1, or EPIC-MOS2), are 
    excluded from the automated fits, and therefore, they are not included in 
    the spectral-fit database.

    \item Complex models (see Sect.~\ref{sec:models}), are only applied if the 
    number of EPIC counts is larger than 500 net counts in the total 0.5--10~keV band.
\end{itemize}

As a result of the application of these criteria, the spectral-fit database contains 
results corresponding to the simple models for 30\,816 source detections, corresponding 
to 22\,677 unique sources (see Table~\ref{tab:sourcenumbers}). Spectral-fitting results 
for complex models are available for 4797 detections. The distribution of spectral 
counts per observation used during the automated fits is plotted in 
Fig.~\ref{fig:xfcz_counthist}. Observations with more than 10\,000 counts,  2\% of the 
30\,816 detections, are excluded of this plot for clarity.

\subsubsection{Photometric redshifts}
\label{sec:photz}
The XMM-\textit{Newton} Photo-Z 
Catalogue\footnote{\url{http://xraygroup.astro.noa.gr/Webpage-prodex/index.html}}
\citep[XMMPZCAT,][]{xmmpzcat} contains photometric redshifts for 100\,178 X-ray 
sources in the 3XMM, that is about 50\% of the X-ray sources within the selected 
XMM-\textit{Newton} observations to build the catalogue. It contains sources outside 
the Galactic plane (|b|>20 deg) with highly reliable optical counterparts in 
the SDSS-DR13 or Pan-STARRS-DR1 catalogues.

This catalogue was produced through a machine learning algorithm, 
MLZ-TPZ \citep{tpz},\footnote{\url{http://matias-ck.com/mlz/}} by using 
optical (SDSS and Pan-STARRS) and (if available) near- (2MASS, UKIDSS, 
VISTA) and/or mid-IR (All-WISE) photometry , to derive photometric redshifts. 
The version of the catalogue used to construct the current XMMPZCAT is 
3XMM-DR6. Optical  and IR data were obtained from multiwavelength cross-matched catalogues using tools and results from the ARCHES project \citep{arches, Pineau17}. 

XMMPZCAT also includes spectroscopic redshifts for 12\,693 sources,
obtained either from the SDSS or other spectroscopic surveys used for
training the machine learning algorithm, like XXL \citep{Menzel16}, 
XMS \citep{xms} or XBS \citep{DellaCeca04}. See \citet{Mountrichas17, xmmpzcat} for a detailed description of these surveys. If available, we used
these spectroscopic redshifts for the X-ray spectral fits.

\begin{table}
\caption{Number of sources}
\centering
\setlength{\tabcolsep}{2.3mm}
\begin{tabular}{lrrr}
\hline
Sample & Detections & Unique & WISE \\
 & & sources & \\
\hline
\hline
XMMPZCAT               &             -        &       100178         &     61998 \\
XMMFITCAT-Z (total) &              30816      &        22677    &          17133 \\
X-ray non-AGN\tablefootmark{a} &    1881    &           1433    &           1181 \\
X-ray AGN unabsorbed  &            17158    &          12686    &          10116 \\
X-ray AGN absorbed    &              977    &            765    &            668 \\
\hline
\label{tab:sourcenumbers}
\end{tabular}
\tablefoot{
\tablefoottext{a}{Sources with a good X-ray spectral modelling but 
with luminosity below our selection criteria for AGN.}
}
\end{table}

\subsection{Bayesian X-ray Analysis}
\label{sec:bxa}
We used the fitting and modelling software {\tt Sherpa 4.9.1} \citep{sherpa} to 
perform the automated spectral fits. We followed the Bayesian technique proposed in 
\citet{Buchner14} using the Bayesian X-ray Analysis (BXA) software, which 
connects the nested sampling algorithm MultiNest \citep{Feroz09} with Sherpa.

By using BXA we can incorporate the whole probability distribution for each
photometric redshift as a prior in the model fitting. Moreover, a nested sampling
algorithm like MultiNest allows for a full exploration of the parameter space, 
avoiding finding only solutions for local minima, a common problem that arise 
when using standard minimization techniques like e.g. the Levenberg-Marquardt 
algorithm.

In the BXA framework, the use of Cash statistics \citep{Cash79} is mandatory.
We used the {\tt wstat} implementation in Sherpa, which allows using background datasets
as background models. {\tt wstat} is used for modelling Poisson data with Poisson 
background when no physically motivated background model is available. It assumes 
a background model with as many parameters as bins in the background spectrum. Under
certain assumptions, there is an analytical solution for calculating the values
of these parameters (see e.g. Appendix B of the Xspec User's
Guide).\footnote{\url{https://heasarc.gsfc.nasa.gov/xanadu/xspec/manual/XSappendixStatistics.html}} For using the {\tt wstat} statistics, the grouped spectra from the 3XMM-DR6
catalogue were ungrouped, and then grouped to 1 count per bin.

All available instruments and exposures for a single observation of a source were 
fitted together, using spectral data within the 0.5--10 keV band. All model
parameters for different instruments are tied together except for a relative 
normalization, which accounts for the differences in flux calibration between
instruments.

\begin{figure}[t]
  \centering
  \resizebox{\hsize}{!}{\includegraphics{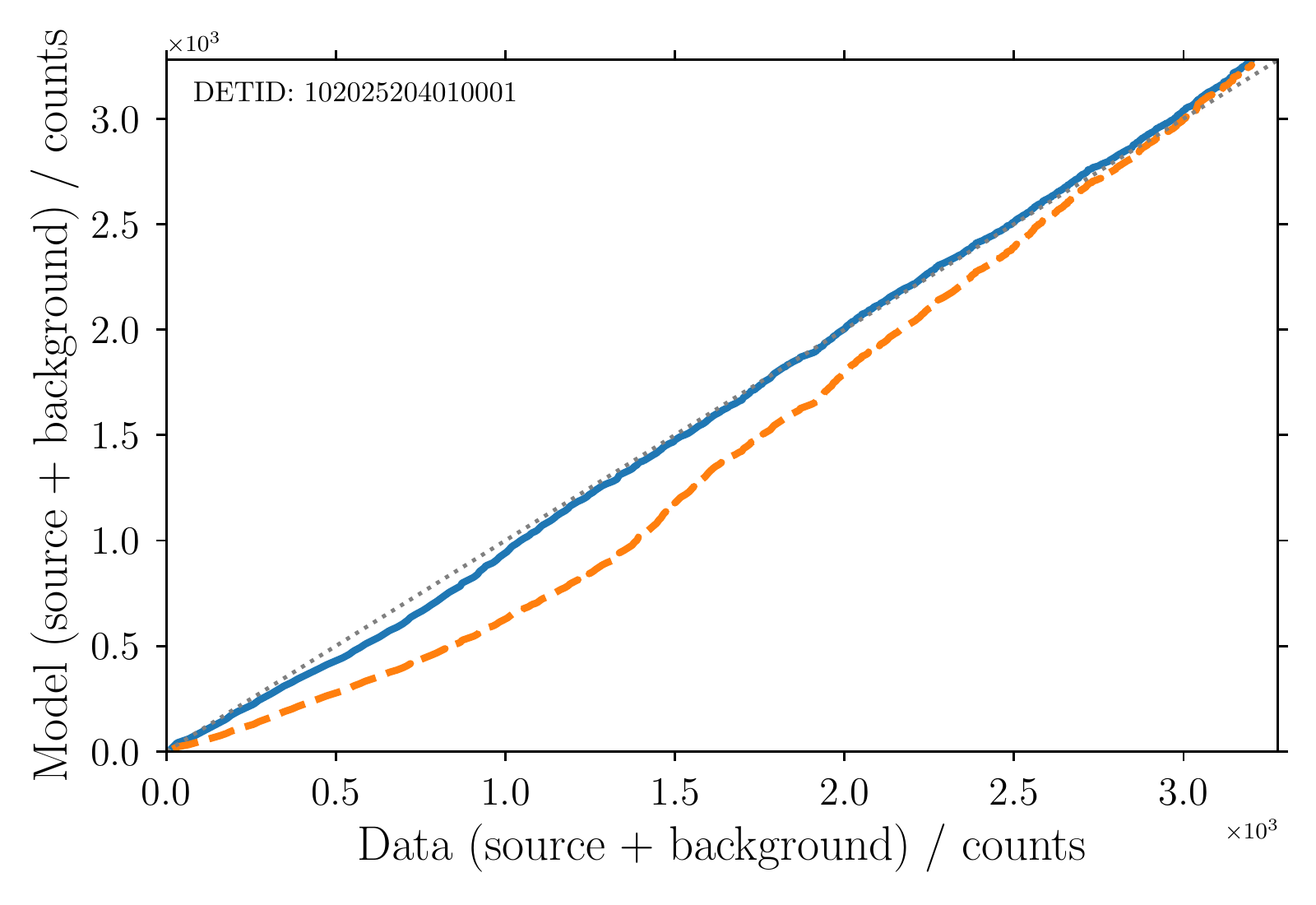}}
  \caption{Q-Q plots for a 3XMM spectrum (EPIC-pn) fitted with a
           an absorbed power-law (blue, solid line) and an 
           absorbed thermal model (orange, dashed line).}
\label{fig:xfcz_qqplots}
\end{figure}

\begin{figure}[t]
  \centering
  \resizebox{\hsize}{!}{\includegraphics{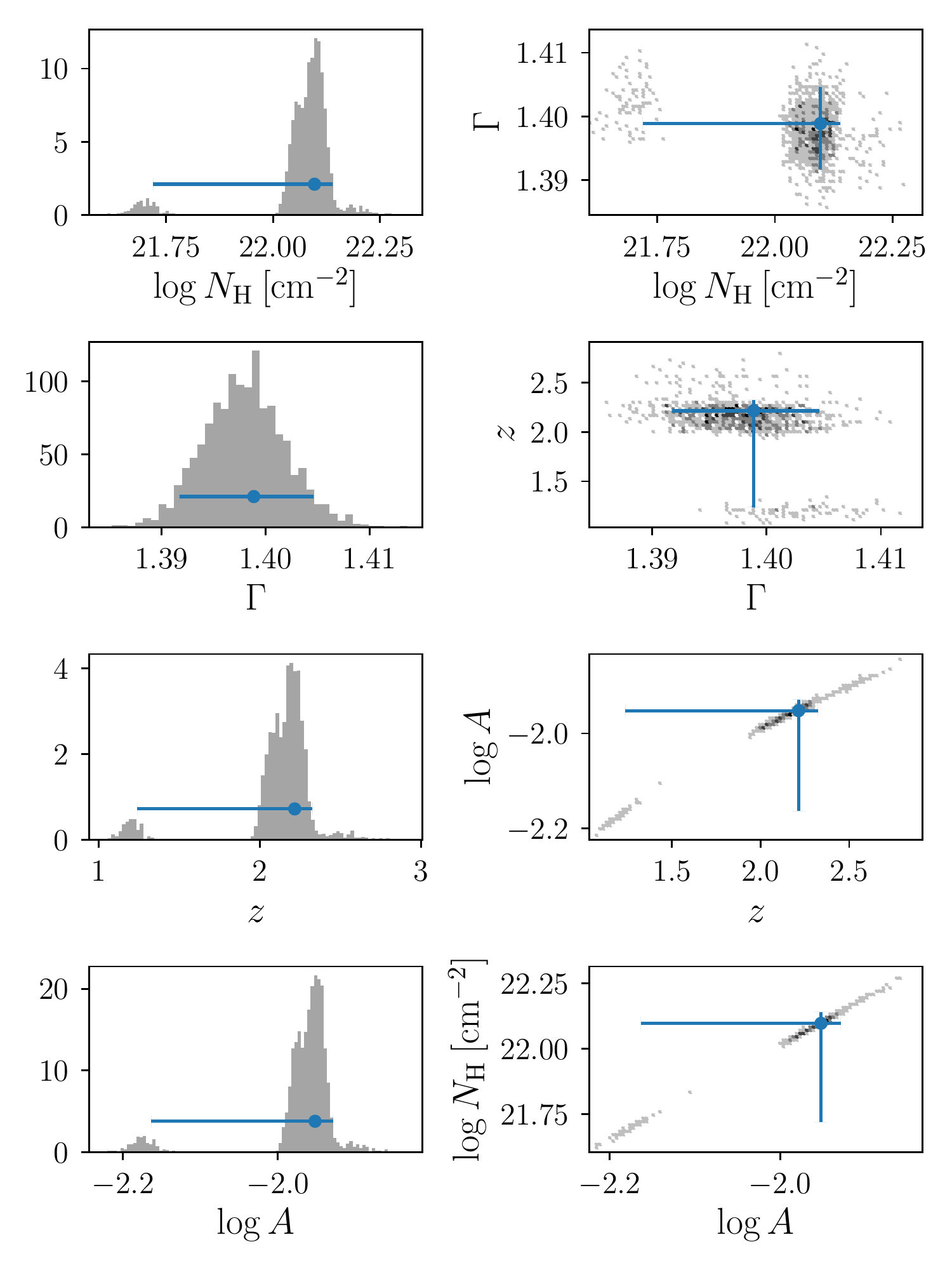}}
  \caption{Posterior distributions for a 3XMM detection (DETID 106909002010001) 
           fitted with an absorbed power-law. \textbf{Left column:}
           marginal distributions for Hydrogen column density, photon index, 
           redshift and power-law normalization. \textbf{Right column:} 
           two-parameters conditional distributions. Blue circles with error
           bars show the best-fit value of the parameters and the corresponding
           90\% credible intervals.}
\label{fig:xfcz_marginals}
\end{figure}

\subsection{Models and priors}
\label{sec:models}
Most sources included in XMMPZCAT are extragalactic and hence the population is 
dominated by AGN. Given this fact, we have reduced the number of spectral models 
with respect to XMMFITCAT. They are phenomenological models selected to reproduced 
the spectral emission of AGN. We also included a simple thermal model to deal 
with the X-ray emission of stars (about ten percent of XMMPZCAT sources) and other 
hot plasmas (e.g. intra-cluster medium emission).

There are four different models, two simple and two more complex models, as follows:
\begin{itemize}
    \item \textbf{Simple models:}
    \begin{itemize}
        \item \textbf{Absorbed power-law model} (``wapo'', 
        {\tt XSPEC: zwabs*pow}): 
        Variable parameters are the Hydrogen column density of the absorber, 
        the power-law photon index, and the power-law normalization.

        \item \textbf{Absorbed thermal model} (``wamekal'', 
        {\tt XSPEC: zwabs*mekal}): 
        Variable parameters are the Hydrogen column density of the absorber, 
        the plasma temperature of the thermal component, and the normalization 
        of the thermal component.
    \end{itemize}

    \item \textbf{Complex models:}
    \begin{itemize}
        \item \textbf{Absorbed thermal plus power-law model} (``wamekalpo'', 
        {\tt XSPEC: zwabs*(mekal + zwabs*pow)}): 
        Variable parameters are the Hydrogen column density of both absorbers, 
        the plasma temperature, the photon index, and the normalization of the
        power-law and thermal components.

        \item \textbf{Absorbed double power-law model} (``wapopo'', 
        {\tt XSPEC: zwabs*(pow + zwabs*pow)}): 
        Variable parameters are the Hydrogen column density of both absorbers,
        the photon indices of both power-law components, and the corresponding  
        normalizations.
    \end{itemize}
\end{itemize}
All models include an additional absorption component, {\tt wabs}, to take into
account the Galactic absorption, with the Hydrogen column density fixed to the
value in the direction of the source from the Leiden/Argentine/Bonn (LAB)
Survey of Galactic HI \citep{nh2}.

In the BXA framework we employed for the spectral fitting, a probability prior 
should be assigned to each free parameter in the model. These are the priors we 
selected:
\begin{itemize}
    \item \textbf{Hydrogen column density:} Jeffreys prior (i.e. a uniform prior 
    in the logarithmic space) with limits $10^{20} - 10^{25}~\mathrm{cm^{-2}}$.

    \item \textbf{Power-law photon index:} Gaussian prior with mean 1.9 and 
    standard deviation 0.15. It corresponds to the photon index distribution 
    for the AGN population as described in \cite{Nandra94}.

    \item \textbf{Thermal plasma temperature:} uniform prior with limits 0.08--20 keV.

    \item \textbf{Redshift:} for sources with known spectroscopic redshift, this 
    is included as a fixed parameter in the models. When only a photometric 
    redshift is available, the redshift is treated as a free parameter using 
    the photo-z probability density distribution calculated in XMMPZCAT as the 
    prior.

    \item \textbf{Normalization:} Jeffreys prior with limits $10^{-30} - 1$.

    \item \textbf{Relative normalization constants:} Jeffreys prior with limits 
    0.01 -- 100.
\end{itemize}

\begin{figure*}[ht]
  \begin{minipage}{0.5\textwidth}
    \centering
    \resizebox{\hsize}{!}{\includegraphics{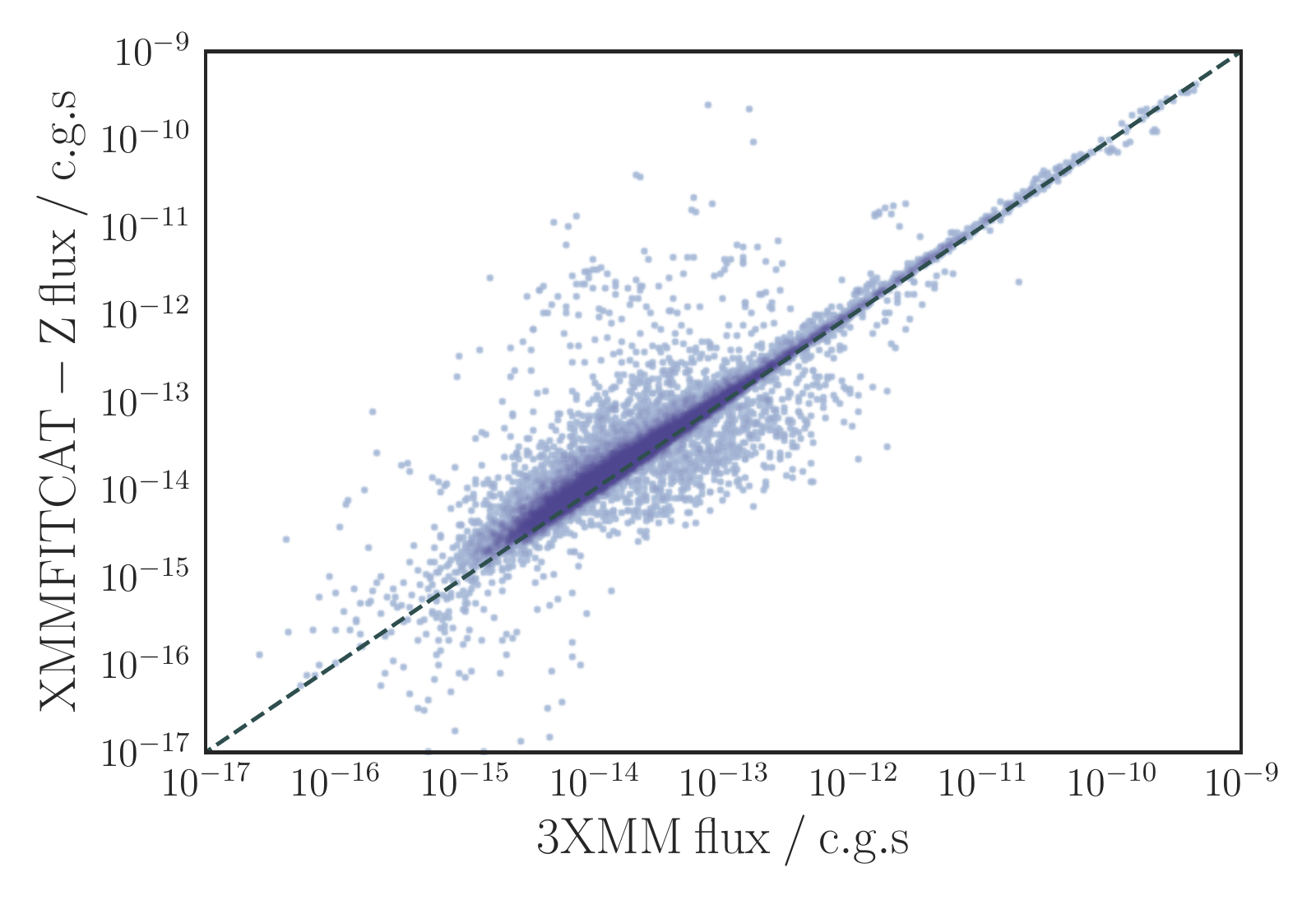}}
  \end{minipage}
  \begin {minipage}{0.5\textwidth}
    \centering
    \resizebox{\hsize}{!}{\includegraphics{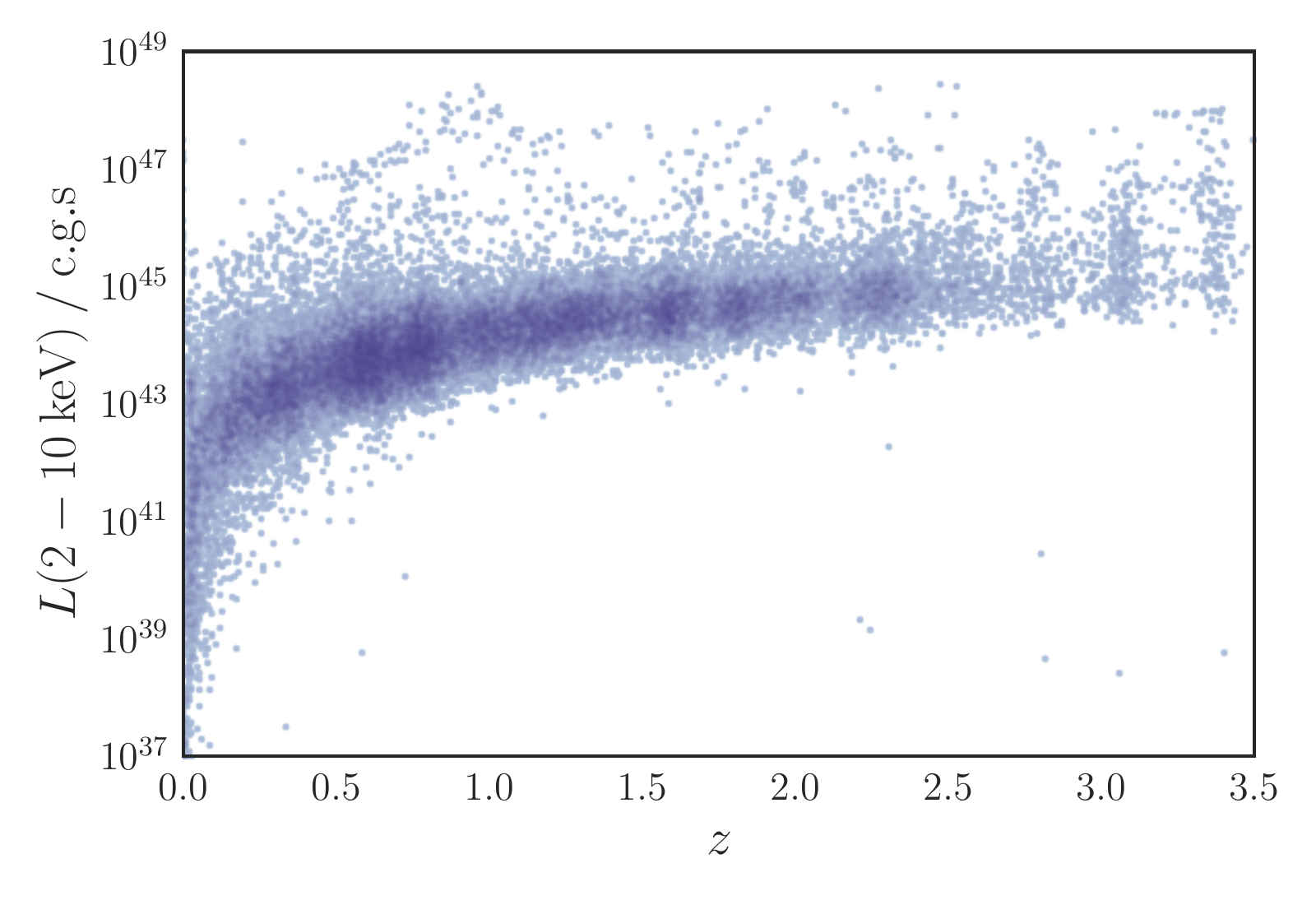}}
  \end{minipage}
   
  \caption[Fluxes and luminosities of XMMFITCAT-Z]{
           \textbf{Left:} Soft fluxes (in c.g.s. units) computed from the 
           automated fits (P\_MODEL) against fluxes in the 3XMM-DR6 catalogue.
           \textbf{Right:} Hard luminosity versus redshift (in c.g.s. units) 
           computed from the automated fits (P\_MODEL). Darker colors show
           a higher density of points.}
\label{fig:xfcz_fluxes}   
\end{figure*}

\subsection{Goodness of fit}
\label{sec:goodness}
Cash maximum likelihood statistics lacks a direct estimate of goodness of fit (GoF). 
We followed the method proposed in \citet{Buchner14} and we used quantile-quantile 
(Q-Q) plots to obtain an estimate of the GoF of our spectral fits.

A Q-Q plot compares the cumulative counts of the data (source + background) with the 
predicted counts (source + background) of the model (see Fig.~\ref{fig:xfcz_qqplots}).
These plots gives a quick visual idea of how well the model can reproduce the data.
For a quantitative estimate of the GoF we calculated the Kolmogorov-Smirnov (KS) 
statistic between the two cumulative distributions, and the corresponding p-value. 
A low KS (or a high p-value) means that the data is well reproduced by the model.

We note however that in this case the p-values for the KS statistic cannot  
be calculated the usual way. The cumulative distribution of the model 
depends on the parameters that were estimated from the data distribution, 
therefore the condition of independence between the two compared 
distributions does not hold, and hence p-values estimated using 
the KS probability distribution are grossly incorrect. Nevertheless, 
through a permutation test we can get an estimate of the p-value. For 
each source, we did 1000 resamplings, randomly splitting the original 
data+model sample in two equal-size subsamples, and estimate the 
corresponding KS statistic. Our estimated p-values are the ratio of 
resamplings that have statistics larger than the statistic of the original 
samples.

Any model showing a KS p-value < 0.01 is considered as an acceptable fit. 
If we found at least one acceptable fit for a given data set, the A\_FIT 
parameter is set to True. The XBA's MultiNest algorithm also estimates the 
evidence for each model ($\log \mathrm{Z}$). The model with the highest
evidence (lowest $\log \mathrm{Z}$) is selected as the preferred model
(P\_MODEL parameter). A preferred model is always included, even in 
the case that it is not an acceptable fit, showing a KS p-value above
0.01. We also list the remaining models (A\_MODELS parameter) with
relative evidence lower than 30, with respect to the preferred model, 
i.e. ``very strong evidence'' accordingly to the scale of \citet{Jeffreys61} 
\citep[see also][]{Robert09,Buchner14}. Assuming that all models are 
\textit{a priori} equally probable, there is no statistical reason to
rule out any of the models in the set formed by P\_MODEL and A\_MODELS.
Without additional information favouring some model, if the fit is
acceptable, the simplest model should be selected as the best-fit model.

We find that 28\,887 detections, 94\% of the XMMFITCAT-Z detections, have an 
acceptable fit. Specifically, 22\,427 detections (73\%) have a single 
absorbed power-law (``wapo'') as its preferred model, and 27\,659 detections 
(90\%) have this model either as preferred or in the accepted models.

\subsection{Parameter and error estimation}
\label{sec:errors}
The MultiNest algorithm implemented in BXA gives the marginalized posterior probability 
distribution for all free parameters in the fitted model. We used these distributions 
to estimate the best-fit parameters and the corresponding errors. The best-fit values 
correspond to the mode (the most probable value) of the posterior distribution, estimated 
using a half-sample algorithm \citep{Robertson74}. Errors were estimated using 
the posterior distributions to calculate a 90\% credible interval around the 
mode.

Figure \ref{fig:xfcz_marginals} shows an example of the marginal and 
conditional (for two parameters) posterior distributions for a source 
fitted using an absorbed power-law. This example shows how the structure 
of the original photo-z probability distribution is preserved.

\subsection{X-ray fluxes and luminosities}
\label{sec:fluxes}
The posterior probability distribution of the free parameters was 
propagated to estimate the flux and errors for each model. This method 
preserves the structure of the uncertainty (degeneracies, multimodal 
structure, etc.). Reported fluxes and luminosities in the catalogue 
correspond to the mode of the posterior distribution, with errors 
estimated as 90\% credible intervals. These are EPIC fluxes, meaning that, 
in the case of multiple instrument spectra for a single observation, the 
reported flux is the average of the different fluxes for each instrument 
and exposure.

For sources with spectroscopic redshifts, luminosities were estimated 
using the intrinsic fluxes and the luminosity distance corresponding to 
that redshift. For sources with photometric redshifts, $z$ is a free 
parameter and hence is propagated with the posterior distribution to 
estimate the corresponding luminosity distance in each case. We assumed 
a $\Lambda$CDM cosmology with $H_0 = 67.7, \Omega_m = 0.307$ \citep{Planck15}.

We estimated fluxes and luminosities in the soft (0.5--2 keV) and 
hard (2--10 keV) bands. For observed fluxes these bands correspond to the 
observer frame. For intrinsic fluxes and luminosities they correspond 
to the source's rest frame.

EPIC soft fluxes obtained from the spectral fits (using results for 
P\_MODEL, see Appendix~\ref{catalogue}) are plotted against those from
the 3XMM-DR6 catalogue (EP\_2\_FLUX + EP\_3\_FLUX) in the left panel of
Fig.~\ref{fig:xfcz_fluxes}, and the hard X-ray luminosity against redshift
in the right panel. Fluxes from the automated fits and from the 3XMM-DR6
are consistent within errors for $\sim 70$\% of the detections. The
conversion from count rates to fluxes in the 3XMM assumes a simple power-law 
model with photon index of 1.7. Hence, we can expect that XMMFITCAT-Z sources 
with different spectral shapes will show larger discrepancies in flux. That 
is indeed the case:  significant differences between flux values are more 
frequent among sources displaying a soft spectrum, that is, sources that are 
best-fitted by a power-law with a steep photon index, or by a thermal model. 
More than 80\% of the non-matching fluxes correspond to any of these cases.

\begin{figure*}[t]
  \begin{minipage}{\textwidth}
    \centering
    \resizebox{\hsize}{!}{\includegraphics{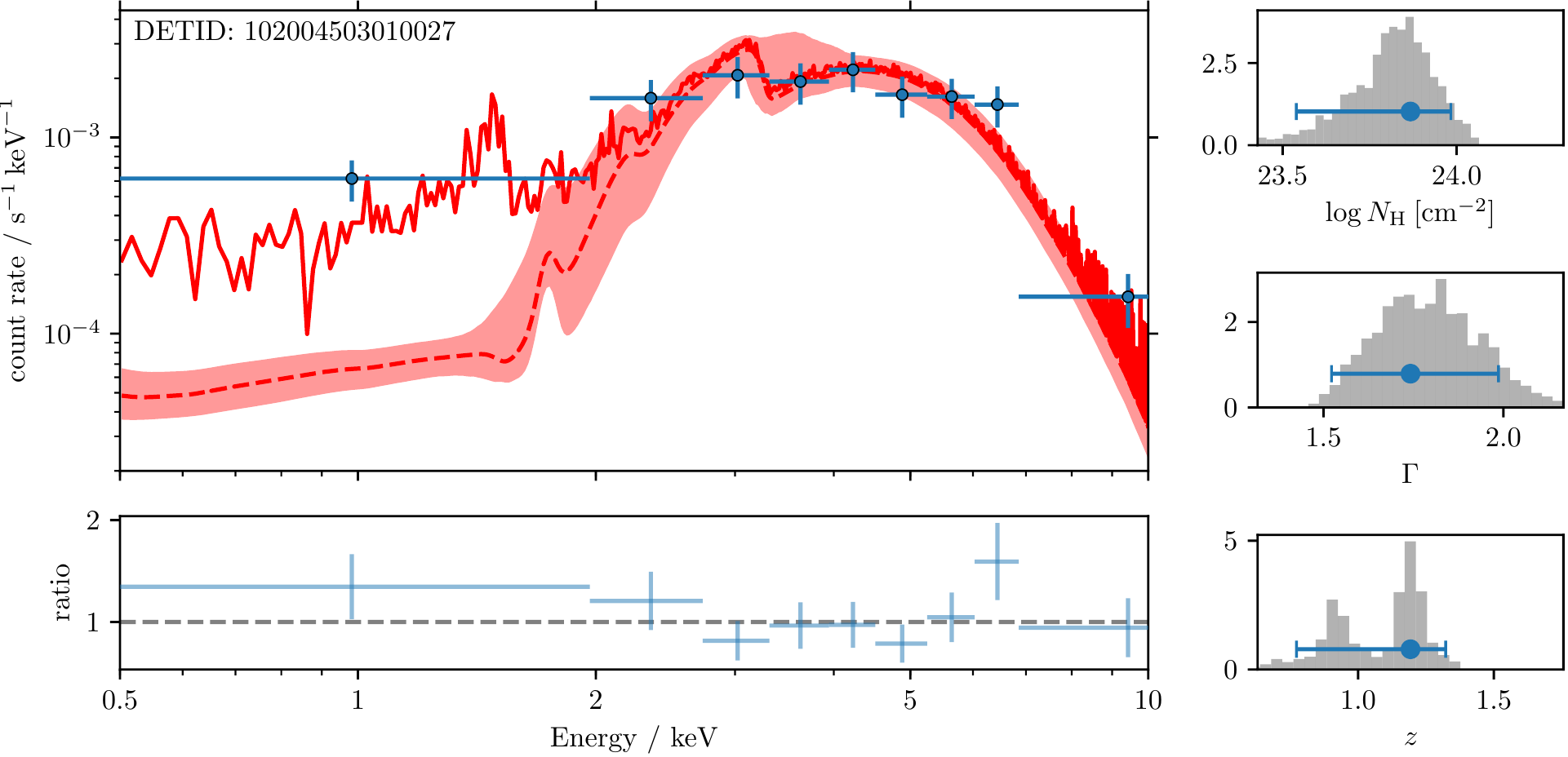}}
  \end{minipage}
  \begin {minipage}{\textwidth}
    \centering
    \resizebox{\hsize}{!}{\includegraphics{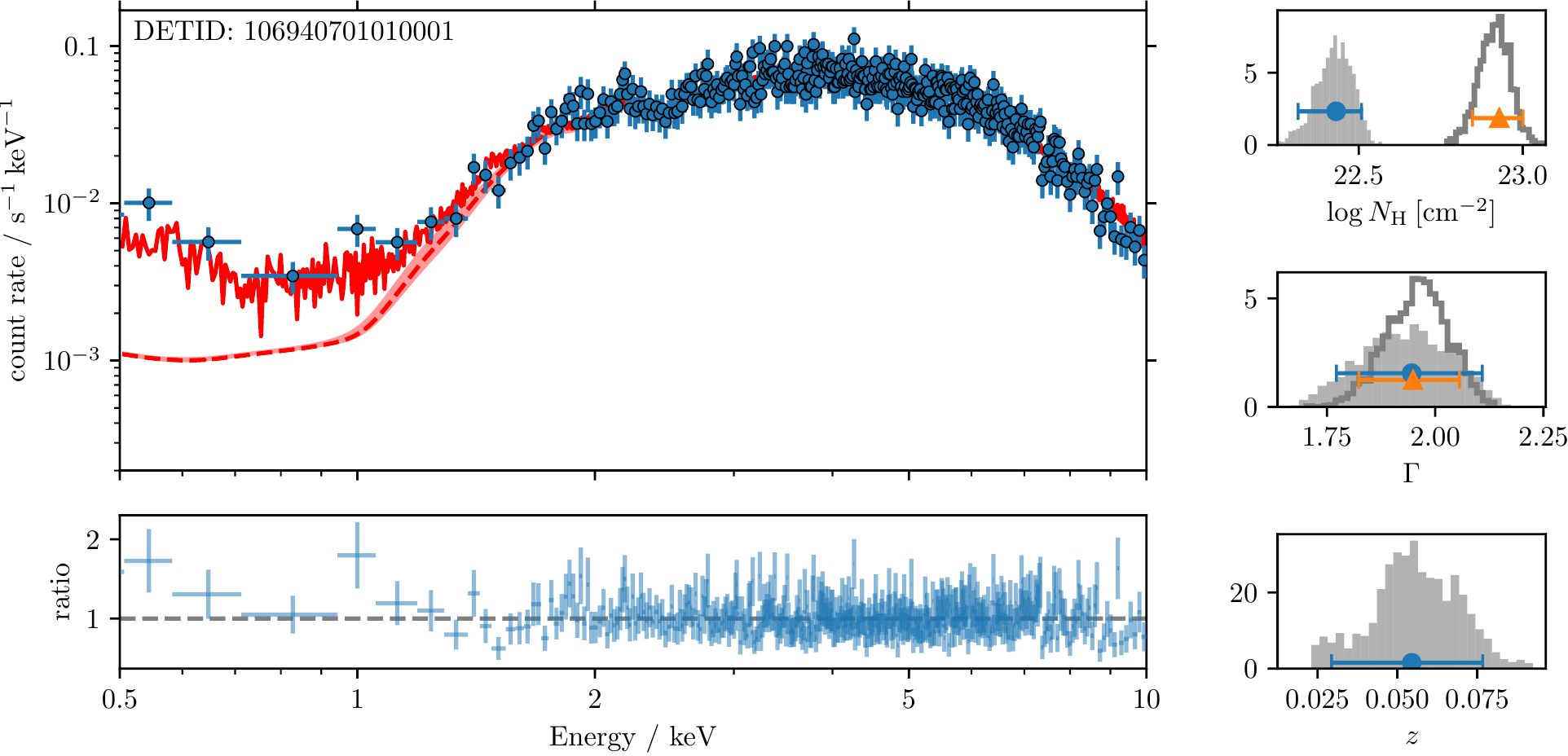}}
  \end{minipage}
   
  \caption{XMM-\textit{Newton} spectra (EPIC-pn) of two X-ray absorbed AGN identified 
          through our selection criteria. For improving visualization, spectral 
          data (blue circles) have been grouped to obtain a signal-to-noise 
          ratio per bin of at least five. Solid red lines: background plus 
          model; dashed, red lines: model; red areas: $2\sigma$ model 
          uncertainty. Panels in the lower part of the plots show the ratio 
          between the observed and the model-predicted count rate. Histograms 
          on the right column show the posterior probability distribution of 
          the model parameters, the corresponding best-fit values (mode) and 
          their 90\% credible intervals. Solid, grey histograms and blue circles
          correspond to the outer absorber and power-law. Open histograms and
          orange triangles correspond to the inner model components.
          \textbf{Top:} detection best fitted with a single absorbed power-law
          (``wapo'' model).
          \textbf{Bottom:} detection best fitted using a double power-law with
          two absorbers (``wapopo'' model).}
\label{fig:specabsorb}
\end{figure*}

\section{Search of X-ray absorbed sources}
\label{sec:xrayabs}
The basic mechanism that explains the panchromatic emission of an AGN is the 
accretion of gas into the central supermassive black hole of the host galaxy. 
The high temperatures in the inner-most regions of the accretion disk ionize 
the surrounding gas, creating a ``hot corona'' of energetic electrons. Optical 
and UV photons interact, through inverse Compton scattering, with the electrons 
in the corona, causing the characteristic X-ray emission of AGN 
\citep{Haardt91,Haardt94}.

This primary emission can be then reprocessed through other physical processes: 
absorption due to ionized/warm/cold material \citep{Turner09}; reflection by 
the material in the accretion disk or by farther neutral gas 
\citep{Fabian06,Turner09}; scattering by an absorbing medium \citep{Netzer98}; 
or fluorescent emission of ionized iron \citep{Mushotzky95,Fabian00}.

The combination of these processes can produce a quite complex X-ray spectrum 
in the observing spectral range of XMM-\textit{Newton} ($\sim0.2-12$~keV). However, this 
complexity is reasonably well captured by the phenomenological models we 
presented in Section~\ref{sec:models}: the X-ray primary emission can be 
modelled by a power-law with photon index $\sim1.9$; Compton Thin neutral 
absorbers ($10^{22} < N_\mathrm{H} <10^{24}$ cm$^{-2}$) are modelled by the 
different absorption components included in our models; Partial absorption can 
be mimicked through double power-law models; Compton Thick absorption 
($N_\mathrm{H} > 10^{24}$ cm$^{-2}$), where all primary emission is suppressed 
below 10 keV, can be also roughly approximated by a double power-law (see 
below); and reflection effects are roughly modelled through power-laws or 
thermal emission components (``wapopo'' or ``wamekalpo'' model). By using these 
simple models we found acceptable fits for 28\,887 out of 30\,816 (94\%) X-ray 
spectra included in XMMFITCAT-Z. 

Our first goal was to identify AGN in XMMFITCAT-Z showing spectral features of
X-ray absorption. Although there exist low-luminosity AGN \citep[see e.g.][]{Ho99, Panessa07},
any persistent astronomical source with an intrinsic (i.e. absorption 
corrected) 2--10 keV luminosity above $10^{42}~\mathrm{erg\, s^{-1}}$ can be 
associated to an AGN, no other known mechanism can produced such luminosity.

Compton Thin absorbed sources can be identify by a significant absorption 
component with $N_\mathrm{H} \gtrsim 10^{22}$ cm$^{-2}$. Compton Thick sources
are harder to identify within the XMM-\textit{Newton} spectral range since, as we stated 
above, all primary emission is suppressed below rest-frame 10 keV. Only the 
X-ray emission scattered by the absorbing medium is observable in this range. 
This emission show a flatter spectral shape with a prominent iron K$_\alpha$
emission line (equivalent width above 500 eV). We must note that the models 
used in XMMFITCAT-Z did not include iron emission lines, and hence Compton Thick 
sources only detectable by this spectral feature could be missed in our final
sample of X-ray absorbed AGN.

\begin{figure}[t]
  \centering
  \resizebox{\hsize}{!}{\includegraphics{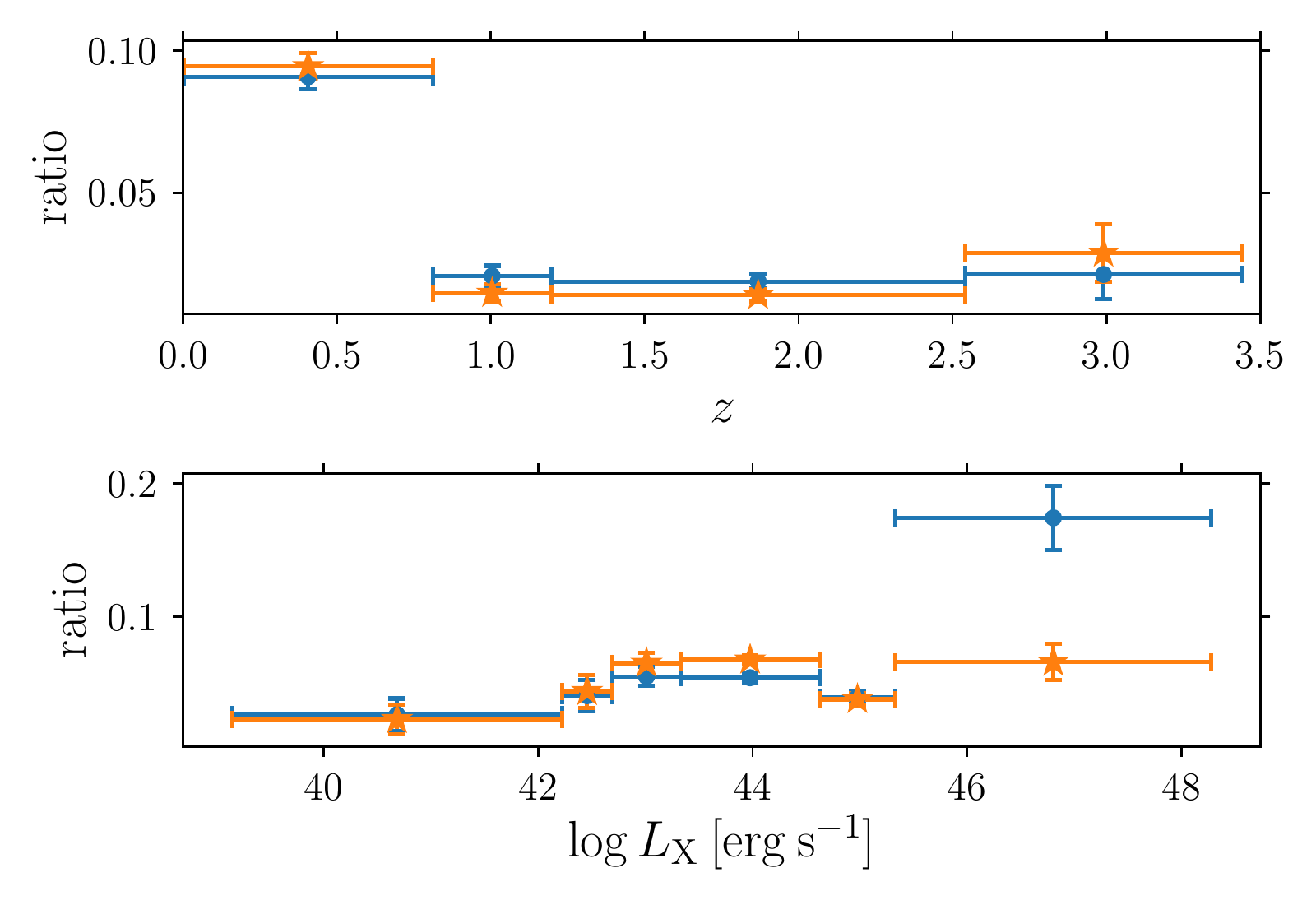}}
  \caption{Ratio between X-ray absorbed and unabsorbed AGN identified in 
           XMMFITCAT-Z for different redshift (top) and luminosity 
           (bottom) ranges. Blue circles correspond to ratios estimated
           using all X-ray absorbed AGN; the ratios shown as orange stars 
           are estimated using only absorbed AGN with a reliable best-fit
           model (see Sect.~\ref{sec:discuss}).}
\label{fig:xabsratio}
\end{figure}

In summary, our selection criteria to identify X-ray absorbed AGN were:
\begin{itemize}
    \item We first selected sources with a single absorbed power-law (``wapo'')
    as an accepted model (P\_MODEL or in A\_MODELS, and A\_FIT true). Those
    sources showing $N_\mathrm{H} > 10^{22}~\mathrm{cm^{-2}}$ or $\Gamma < 1.4$,
    with 90\% confidence,\footnote{The lower limit of the 90\% credible interval
    for the parameter is above our selection limit.} are classified as absorbed.
    
    \item Among those sources without ``wapo'' as an accepted model, we select
    those with a double power-law with two absorbers (``wapopo'') as an accepted
    model. Compton Thin sources can be identified as those showing $N_\mathrm{H} > 10^{22}~\mathrm{cm^{-2}}$ in the inner absorber (logNH2). This model also
    allows for a crude approximation of the spectrum of Compton Thick objects:
    those showing a high obscuration in the inner absorber
    ($N_\mathrm{H} > 10^{23}~\mathrm{cm^{-2}}$) and $\Gamma < 1.4$ in the outer
    power-law (PhoIndex2).

    \item Finally, among those sources classified as absorbed, we selected only 
    those showing an intrinsic 2--10~keV luminosity consistent with being greater
     than $10^{42}~\mathrm{erg\, s^{-1}}$ (i.e. upper error bars are above $10^{42}$).
\end{itemize}

Applying these selection criteria we found 1421 spectra with signs of X-ray 
absorption ($\sim 5\%$ of the total catalogue), corresponding to 1037 unique
sources (out of 22\,677 unique sources in the total catalogue); 1385 out of 1421
show $N_\mathrm{H} > 10^{22}~\mathrm{cm^{-2}}$ and 38 show a flat spectrum with 
$\Gamma < 1.4$. Two of them show both high $N_\mathrm{H}$ and a flat emission.

\begin{figure}[t]
  \centering
  \resizebox{\hsize}{!}{\includegraphics{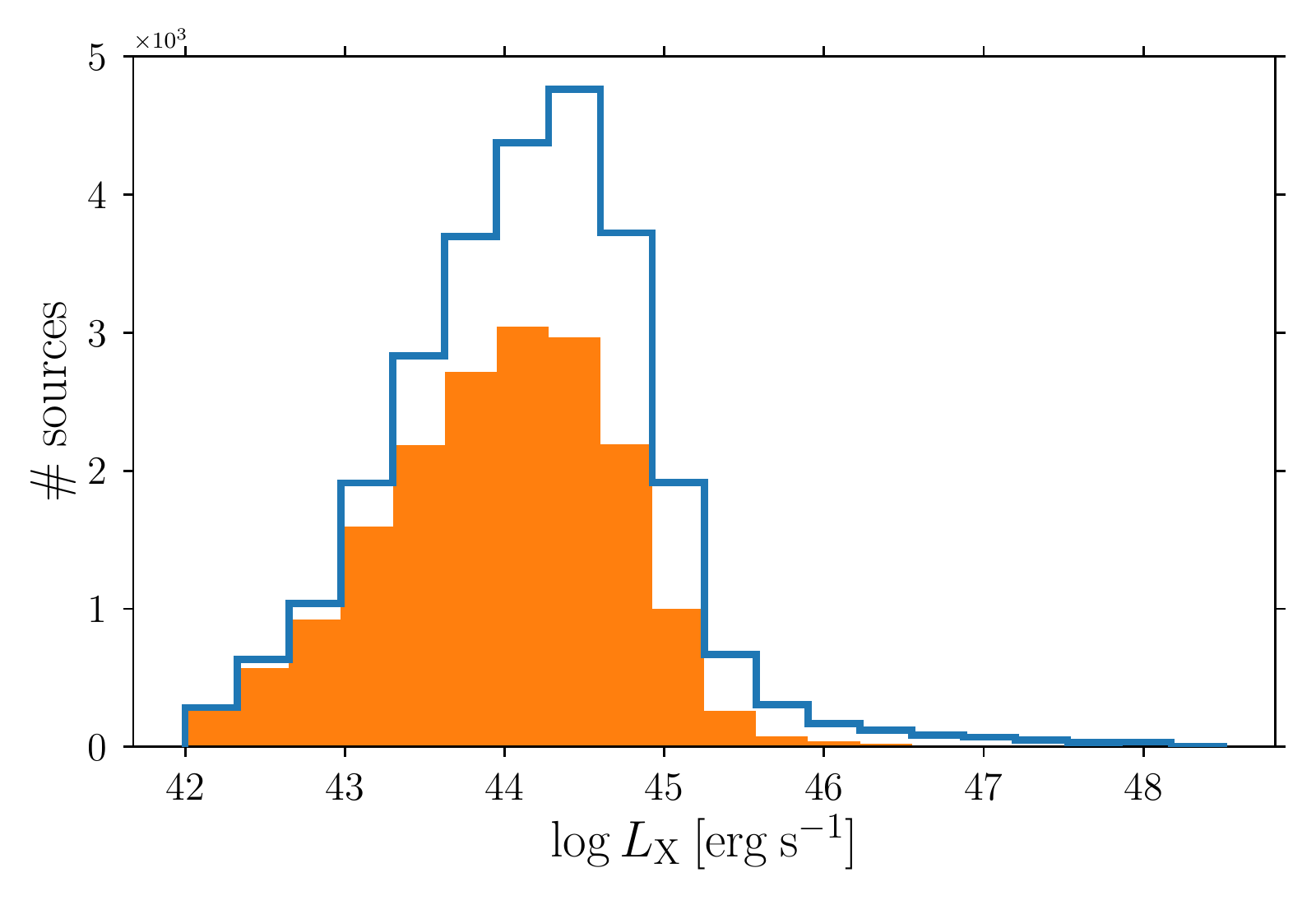}}
  \caption{X-ray luminosity distribution for XMMFITCAT-Z sources classified as 
           AGN. The solid blue line shows the whole sample without any additional
           filtering. The solid orange histogram includes only AGN with a reliable
           best-fit model (see Sect.~\ref{sec:discuss}).}
\label{fig:agnlxhist}
\end{figure}

Figure~\ref{fig:specabsorb} show two examples of 3XMM detections 
included in XMMFITCAT-Z and classified as absorbed AGN following these 
criteria. The plots show their EPIC-pn spectra, including the best-fit model, 
and the corresponding best-fit parameters and their probability distributions.

To build our final selection of  X-ray absorbed AGN we examined further
the X-ray properties of XMMFITCAT-Z AGN (see Sect.~\ref{sec:discuss}). We 
found that our X-ray modelling for a significant fraction of sources
overestimates their X-ray luminosity. This was caused by either a wrong
estimation of their photometric redshifts or due to modelling artifacts
introduced when using a double power-law model (``wapopo''). As we explain
in Sect.~\ref{sec:discuss} below, after applying an additional filtering
for selecting objects with more reliable redshifts and X-ray modelling,
our atlas of X-ray absorbed AGN contains 977 detections, corresponding
to 765 unique sources. We provide this sample as a FITS table.\footnote{\url{http://xraygroup.astro.noa.gr/Webpage-prodex/bin/XMMFITCATZ_3XMMDR6_v1.0_absorbed_agn.fits}}

\section{Discussion}
\label{sec:discuss}
In Fig.~\ref{fig:xabsratio} we present the ratio of absorbed to unabsorbed AGN
as a function of redshift (top panel) and X-ray luminosity (lower panel).
Redshift and luminosity bins were estimated using the Bayesian Block Algorithm
\citep{Scargle13}. This algorithm uses a fitness function calculated on the actual
data for estimating an optimal binning. It allows varying bin sizes for a better reproduction of the underlying distribution. Blue circles show the results
applying the absorption criteria we presented in Sect.~\ref{fig:specabsorb}
based solely in the X-ray spectral fits. 

\begin{figure*}[t]
  \begin{minipage}{0.5\textwidth}
    \centering
    \resizebox{\hsize}{!}{\includegraphics{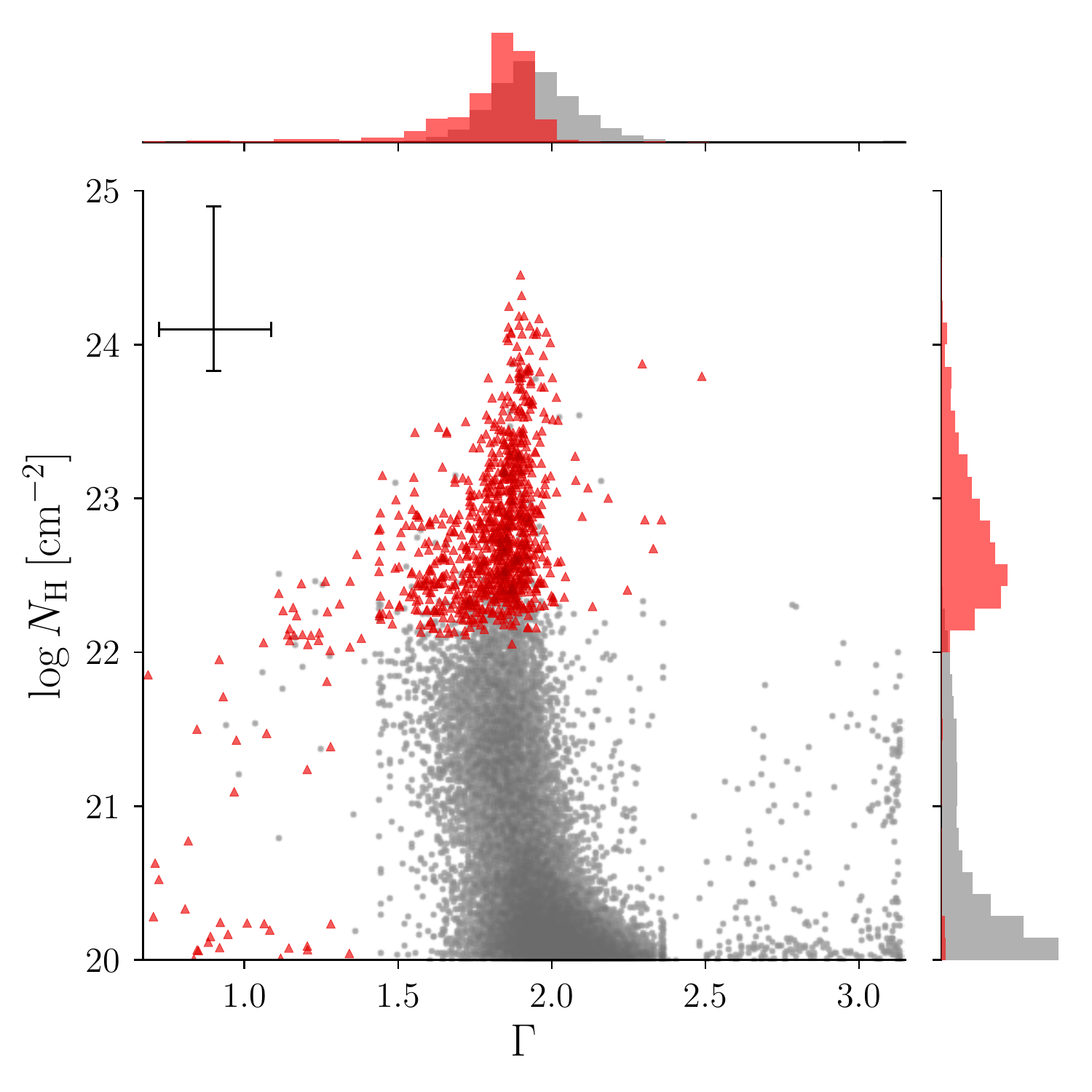}}
  \end{minipage}
  \begin{minipage}{0.5\textwidth}
    \centering
    \resizebox{\hsize}{!}{\includegraphics{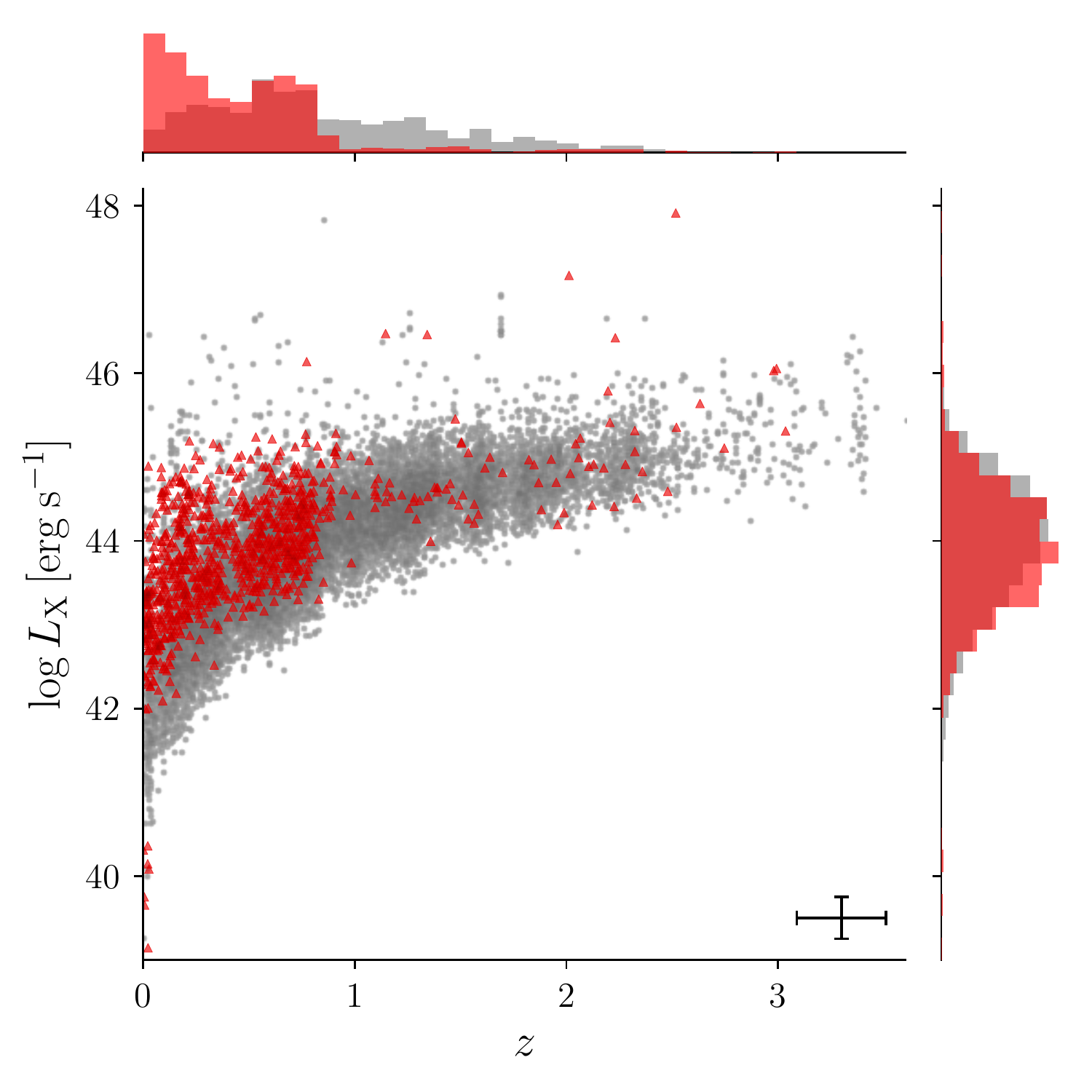}}
  \end{minipage}
  \caption{Photon index versus Hydrogen column density (left) and redshift 
           versus 2--10 keV, absorption corrected luminosity (right) for 
           detections classified as AGN with a reliable X-ray modelling
           (see Sect.~\ref{sec:discuss}). Red triangles are X-ray absorbed
           AGN; unabsorbed AGN are shown as grey circles.
           Errors are not included in the plots to improve visualization. 
           Instead, we include error bars in the corners showing the median
           error of the corresponding parameters. The histograms in the upper 
           and right side of the plots show the normalize distribution of the
           parameters for X-ray absorbed (red) and unabsorbed (grey).}
\label{fig:agnparams}
\end{figure*}

It appears that there is an excess of absorbed sources at low redshift (see also
Fig.~\ref{fig:agnparams}). This is easily explained as the absorbed sources have
a lower flux and thus they are detected at closer distances. The fraction of
absorbed AGN remains roughly constant with luminosity, except for the highest
luminosity bin, where the number of absorbed sources increases significantly.
Although we emphasize that our sample is not statistically complete, but in
contrast biased towards the brightest sources, regarding  both the X-ray and 
the optical selection, we should not expect a strong correlation for the 
fraction of absorbed AGN with X-ray luminosity \citep[see e.g.][]{Sazonov15}.

The blue solid line in Fig.~\ref{fig:agnlxhist} shows the 2--10 keV luminosity
distribution for all XMMFITCAT-Z sources classified as AGN, again based solely
in their X-ray properties. There is a non-negligible high luminosity long tail
for these sources, reaching X-ray luminosities above $10^{47}~\mathrm{erg\,
s^{-1}}$. Such high luminosities are clearly nonphysical.

These results strongly suggest that our X-ray spectral modelling is overestimating
the intrinsic 2--10 keV luminosity for some objects. We therefore tried to 
identify those sources with unreliable modelling. 

There are two main effects that can overestimate the X-ray luminosity: On
one hand, the photometric redshifts calculated for some sources could be
wrong. For building XMMFITCAT-Z we did not apply any filtering regarding
the quality of the photometric redshifts. In \citet{xmmpzcat} we presented
several quality diagnostics for photo-z based on the shape of their
probability density function. For example, by using the Peak Strength (PS)
parameter included in XMMPZCAT we can select sources which photo-z probability
distribution is narrowly concentrated around a single redshift, avoiding
highly multimodal distributions. We considered objects with PS < 0.7 as
having an unreliable X-ray spectral modelling.

On the other hand, the complex models we used for sources with high counts
(the absorbed double power-law model, ``wapopo'', or the absorbed thermal 
plus power-law  model, ``wamekalpo'', see Sect.~\ref{sec:models}) can introduce
artifacts in the X-ray modelling leading to nonphysical parameters: some of
this sources are not well fitted by a single power low or thermal emission 
because they have a small but significant excess emission at the high energy
end of the spectrum. This excess, that is associated with the AGN Compton 
reflection hump \citep{pexrav,Reynolds99}, can be easily modelled by 
introducing a highly absorbed power-law, with $N_\mathrm{H} \gtrsim 10^{23}
\mathrm{cm^{-2}}$. When such artifact appears in the selected best-fit model,
it can clearly overestimate the intrinsic 2--10 keV luminosity. Our selection 
criteria for X-ray absorbed sources will identify this kind of sources as 
highly absorbed, misinterpreting it as a heavily buried AGN component, but in
fact they do not show any real spectral feature for X-ray absorption. 

Hence, we classified as unreliable those X-ray sources well modelled using
any of our complex models and classified as absorbed, but showing a 2--10 keV
luminosity above $10^{45.3}~\mathrm{erg\, s^{-1}}$. This luminosity limit
corresponds to the last luminosity bin in Fig.~\ref{fig:xabsratio}.

Finally, we also excluded sources with a loosely constrained $N_\mathrm{H}$
(i.e. the 90\% credible interval for $\log N_\mathrm{H}$ is greater than 2 dex).
Once we applied these quality criteria for selecting sources with a reliable
X-ray modelling, we rejected 8871 detections: 444 absorbed AGN and 8427
unabsorbed AGN. Orange stars in Fig.~\ref{fig:xabsratio} and the orange solid
histogram in Fig.~\ref{fig:agnlxhist} include only AGN with reliable X-ray
modelling. The figures clearly show that the effects due to overestimated
luminosities disappear after applying this filtering. 

In Fig.~\ref{fig:agnparams} we plot the Hydrogen column density against the
Photon Index (left) and redshift against 2--10 keV luminosity (right) for
detections classified as AGN in XMMFITCAT-Z (objects with unreliable X-ray
modelling are not included), and the corresponding normalized distribution
for each parameter. Grey circles correspond to unabsorbed objects (17\,158 
detections); X-ray absorbed objects are plotted as red triangles (977 
detections).

\begin{figure}[t]
  \centering
  \resizebox{\hsize}{!}{\includegraphics{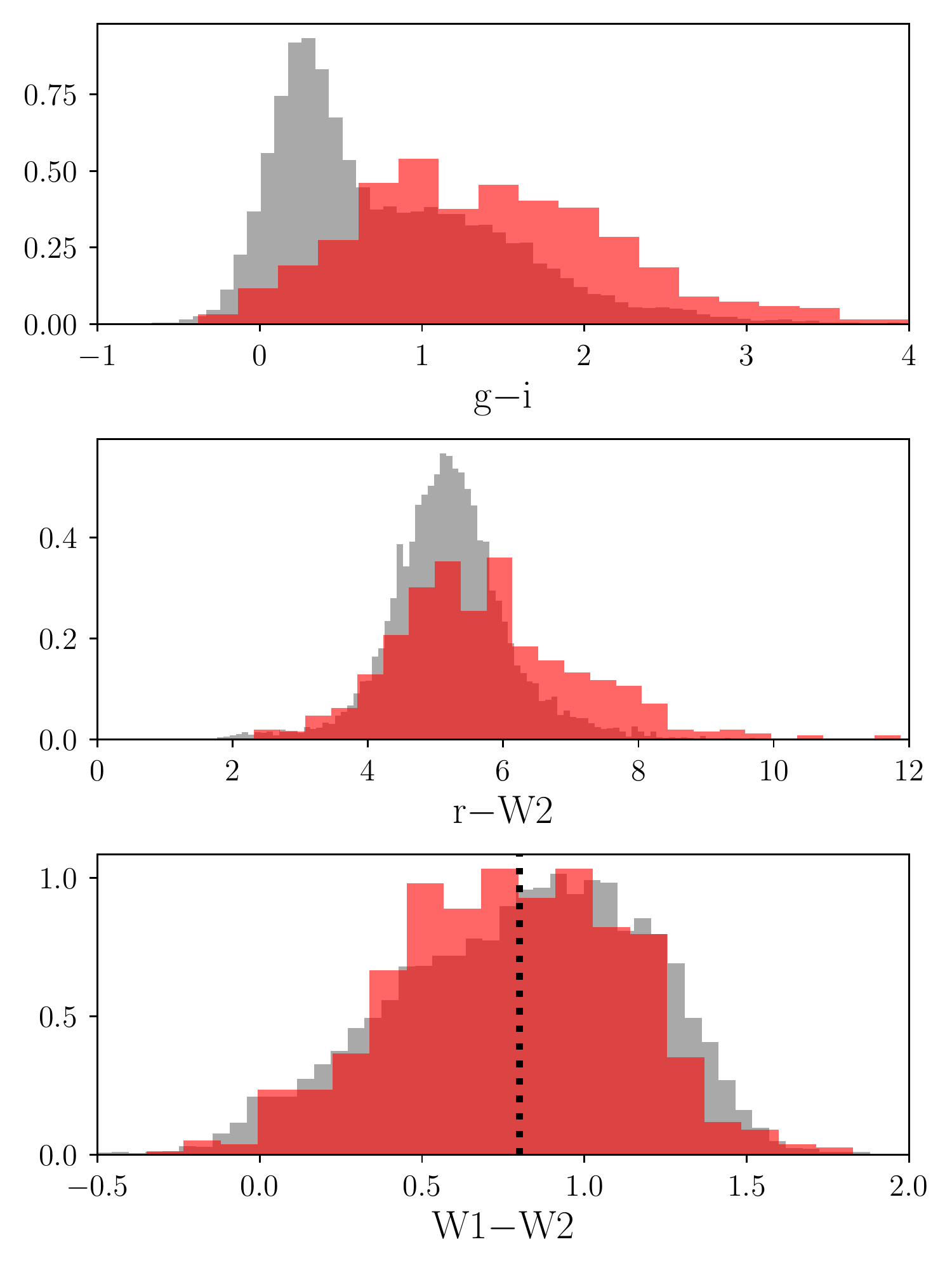}}
  \caption{g--i (top), r--W2 (middle) and W1--W2 (bottom) normalized colour
           distributions. Solid, red histograms: unique sources classified as 
           X-ray absorbed AGN in XMMFITCAT-Z; solid, grey histogram: X-ray
           unabsorbed AGN. Only AGN with reliable modelling are included
           (see discussion in Sect.~\ref{sec:discuss}).
           The vertical dotted line in the bottom panel shows the \citet{Stern12}
           MIR criterion for AGN selection ($\mathrm{W1-W2} > 0.8$).}
\label{fig:histcolors}
\end{figure}

The use of MIR colours is a common tool for AGN identification \citep{Stern05,Donley07,Donley12,Mateos12,Assef13}, in particular by using 
data in the W1 ($3.4\:\mu m$) and W2 ($4.6\:\mu m$) bands from the WISE all-sky 
surveys \citep{wise,neowise,allwise}, since galaxies are expected to display 
bluer colors than AGN in the MIR. In Fig.~\ref{fig:histcolors} we present
the W1--W2 distribution (lower panel) for XMMFITCAT-Z AGN with counterparts in
the All-WISE catalogue \citep{allwise}.

The W1--W2 colour has been proposed by \citet{Stern05} as an efficient diagnostic
for the selection of AGN. The idea is that the heated torus dust results in 
increased emission in the W2 band. MIR colors show no difference between absorbed
and unabsorbed objects, as expected if the MIR emission is from the torus and 
hence inclination independent. We see that a considerable fraction of our sources
is classified as AGN following the W1--W2 criterion. However, an almost equally
large number of X-ray sources would not be classified as AGN following the W1--W2
criterion.

\begin{figure}[t]
  \centering
  \resizebox{\hsize}{!}{\includegraphics{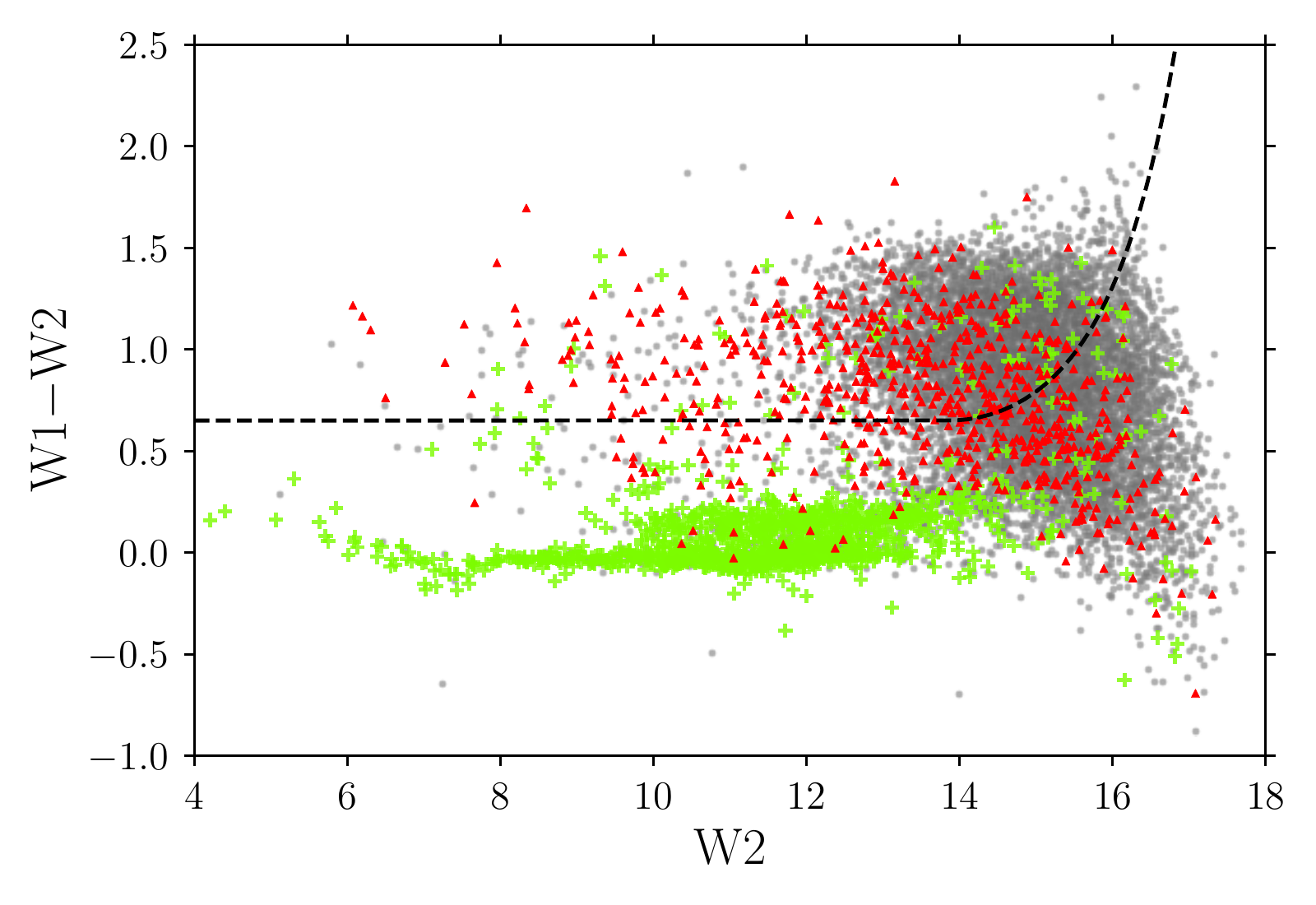}}
  \caption{W2 versus W1--W2 for XMMFITCAT-Z unique sources with WISE
           counterparts. Red triangles show X-ray absorbed AGN; Grey
           circles show X-ray unabsorbed AGN. Light green crosses are 
           objects not classified as AGN, but with an acceptable fit 
           in XMMFITCAT-Z. Only sources with reliable modelling are
           included (see discussion in Sect.~\ref{sec:discuss}). 
           The black, dashed line show the \citet{Assef13} criterion
           for AGN selection.}
\label{fig:mircolors}
\end{figure}

These objects are associated with AGN where the galaxy starts to dominate over 
the AGN colours \citep[see also][]{Barmby06}. \citet{Assef13} propose a more
elaborate criterion where the W1--W2 cutoff is a function of the W2 magnitude.
This selection curve (90\% completeness) is shown in Fig.~\ref{fig:mircolors}.
It is evident that a significant number of X-ray selected AGN would evade
identification if we used mid-IR colour criteria alone. In the same figure
we plot the sources which were not classified as AGN in our spectral fits
(i.e. objects having a good X-ray spectral fit with at least one of our
models, but with 2--10~keV luminosity below $10^{42}~\mathrm{erg\, s^{-1}}$).
The vast majority of these lie outside the AGN locus. This suggests that
the X-ray spectral criteria alone are quite efficient in selecting AGN. 
As stated above, AGN selections based on WISE colours are efficient for
identifying sources where the mid-IR torus emission dominates over the 
host galaxy contribution, that is, for luminous, unobscured AGN 
\citep[see e.g.][]{Eckart10, Hickox17, Pouliasis20}. Our results show
how X-ray spectral criteria can select AGN with lower luminosities. 
Figure~\ref{fig:mircolors} also shows some contamination of sources
within the AGN wedge but with no evidence in X-rays of hosting an AGN 
(light green crosses). \citet{Georgakakis20} suggested that these sources
can be star forming galaxies at redshift $\lesssim 0.5$ that scatter into the 
AGN MIR wedge due to the photometric uncertainties of the WISE magnitudes.

Next, we discuss the obscuration properties of our sample. In the top panel of 
Fig.~\ref{fig:histcolors} we present the g--i colours of the absorbed vs. the 
unabsorbed sources. X-ray absorbed sources are optically redder compared to
unabsorbed sources, as expected. However, there is a non-negligible number
of X-ray unabsorbed sources with very red colours. Part of them could be
missed absorbed objects, but they could be also truly peculiar sources. The
catalogue is then quite useful to identify this kind of objects for further
studies.

In the middle panel of Fig.~\ref{fig:histcolors} we present the r--W2 colour
distribution. \citet{Yan13} suggested that this colour (r--W2 > 6) offers
a powerful diagnostic for the selection of obscured AGN, in combination with
pure MIR diagnostics to identify AGN (W1--W2 > 0.8 and W2 < 15.8) This is because
in the presence of reddening, the r band is easily diminished while the W2
emission remains relatively unscathed. As expected the X-ray obscured sources
have a redder tail compared to the unobscured sources. It is worth noting
however, that a considerable fraction of X-ray unabsorbed sources have
r--W2 > 6. \citet{Hickox17} show that the r--W2 criterion is only really 
effective above $z>1$. The fact that the bulk of our absorbed AGN is below 
redshift one (see Fig.~\ref{fig:agnparams}, right panel) can explain our result.

Our analysis of XMMFITCAT-Z sources is consistent with many previous studies \citep{Eckart10,Stern12,Assef13,Yan13,Mateos12,Mateos13,Mountrichas17b,Mountrichas20,Pouliasis20} showing that MIR diagnostics are powerful for the identification of luminous 
AGN, but they fail for AGN with lower luminosity and/or mild absorption, where
the stellar contribution to the optical/MIR emission is non-negligible. By 
focusing on the X-ray properties, we can identify these objects.

\section{Summary}
\label{sec:conclusion}
We present a catalogue of automated X-ray spectral fits, using the 3XMM-DR6 
spectral products. Photometric redshifts have been previously estimated 
\citep{xmmpzcat} using machine learning techniques \citep[TPZ;][]{tpz}. 
22677 X-ray sources are presented with a reasonable quality X-ray spectrum,
meaning that each XMM-\textit{Newton} detector have over 50 counts, and with 
available photometric or spectroscopic redshift. The X-ray spectral fits have 
been performed using Bayesian X-ray analysis \citep{Buchner14}, including the
PDF of photometric redshift as priors.

As a brief demonstration of the potential of this X-ray spectral catalogue, 
we present the properties of the 765 sources showing strong evidence of being X-ray absorbed AGN. We show that a considerable fraction of our sample would not be
classified as AGN based on their mid-IR colours, W1--W2 vs. W2. About one third
of the X-ray absorbed AGN presents red $\rm r-W2>6$ colours. It appears then
that the r--W2 criterion, often used in the literature for the selection of
obscured AGN, produces very different samples compared to the X-ray criteria
based on the hydrogen column density. 



\begin{acknowledgements}
This work is part of the Enhanced XMM-\textit{Newton} Spectral-fit Database project, 
funded by the European Space Agency (ESA) under the PRODEX program. 
AR acknowledges support of this work by the PROTEAS II project (MIS 5002515), which 
is implemented under the ``Reinforcement of the Research and Innovation Infrastructure'' 
action, funded by the ``Competitiveness, Entrepreneurship and Innovation'' operational 
programme (NSRF 2014-2020) and co-financed by Greece and the European Union (European 
Regional Development Fund).
AC acknowledges financial support from the Spanish Ministry MCIU under project 
RTI2018-096686-B-C21 (MCIU/AEI/FEDER/UE), cofunded by FEDER funds and from the 
Agencia Estatal de Investigación, Unidad de Excelencia María de Maeztu, ref. 
MDM-2017-0765. 
\\
This research has made use of data obtained from the 3XMM XMM-\textit{Newton} 
serendipitous source catalogue compiled by the 10 institutes of the XMM-\textit{Newton} 
Survey Science Centre selected by ESA.
\\
This work is based on observations made with XMM-\textit{Newton}, an ESA science 
mission with instruments and contributions directly funded by ESA Member States 
and NASA. 
\\
Funding for the Sloan Digital Sky Survey IV has been provided by the Alfred P. 
Sloan Foundation, the U.S. Department of Energy Office of Science, and the 
Participating Institutions. SDSS-IV acknowledges support and resources from the 
Center for High-Performance Computing at the University of Utah. The SDSS web 
site is \url{www.sdss.org}.
\\
SDSS-IV is managed by the Astrophysical Research Consortium for the 
Participating Institutions of the SDSS Collaboration including the Brazilian 
Participation Group, the Carnegie Institution for Science, Carnegie Mellon 
University, the Chilean Participation Group, the French Participation Group, 
Harvard-Smithsonian Center for Astrophysics, Instituto de Astrof\'isica de 
Canarias, The Johns Hopkins University, Kavli Institute for the Physics and 
Mathematics of the Universe (IPMU) / University of Tokyo, Lawrence Berkeley 
National Laboratory, Leibniz Institut f\"ur Astrophysik Potsdam (AIP),  
Max-Planck-Institut f\"ur Astronomie (MPIA Heidelberg), Max-Planck-Institut 
f\"ur Astrophysik (MPA Garching), Max-Planck-Institut f\"ur Extraterrestrische 
Physik (MPE), National Astronomical Observatories of China, New Mexico State 
University, New York University, University of Notre Dame, Observat\'ario 
Nacional / MCTI, The Ohio State University, Pennsylvania State University, 
Shanghai Astronomical Observatory, United Kingdom Participation Group, 
Universidad Nacional Aut\'onoma de M\'exico, University of Arizona, University 
of Colorado Boulder, University of Oxford, University of Portsmouth, University 
of Utah, University of Virginia, University of Washington, University of 
Wisconsin, Vanderbilt University, and Yale University.
\\
This publication makes use of data products from the Wide-field Infrared 
Survey Explorer, which is a joint project of the University of California, Los 
Angeles, and the Jet Propulsion Laboratory/California Institute of Technology, 
funded by the National Aeronautics and Space Administration.
\\
This research made use of Astropy, a community-developed core Python package 
for Astronomy \citep{astropy2}.
\end{acknowledgements}

\bibliographystyle{aa}
\bibliography{astrobib}

\begin{appendix}
\section{Description of the XMMFITCAT-Z}
\label{catalogue}
The XMMFITCAT-Z table contains one row for each detection, and 157 columns containing 
information about the source detection and the spectral-fitting results. Not available 
values are represented by an empty "NULL" value. The first 14 columns contain 
information about the source and observation, including redshift information, whereas 
the remaining 143 columns contain, for each model applied, spectral-fit flags, 
parameter values and errors, fluxes, luminosities, and five columns to describe the 
goodness of the fit.

\begin{enumerate}
    \item \textbf{Source and observation}
    \begin{itemize}\setlength\itemsep{0.2em}
        \item IAUNAME: The IAU name assigned to an unique source in the 
        3XMM-DR6 catalogue.

        \item SC\_RA, SC\_DEC: Right ascension and declination in degrees 
        (J2000) of the unique source, as in the 3XMM-DR6 catalogue. RA and 
        DEC correspond to the SC\_RA and SC\_DEC columns in the 3XMM-DR6 
        catalogue. These are corrected source coordinates and, in the case 
        of multiple detections of the same source, they correspond to the 
        weighted mean of the coordinates for the individual detections.

        \item SRCID: A unique number assigned to a group of catalogue 
        entries which are assumed to be the same source in 3XMM-DR6.

        \item DETID: A consecutive number which identifies each entry 
        (detection) in the 3XMM-DR6 catalogue.

        \item OBS\_ID: The XMM-\textit{Newton} observation identification, as in 
        3XMM-DR6.

        \item SRC\_NUM: The (decimal) source number in the individual 
        source list for this observation (OBS\_ID), as in 3XMM-DR6.
        In the pipeline products this number is used in hexadecimal 
        form.

        \item PHOT\_Z, PHOT\_ZERR: Photometric redshift of the source (from 
        XMMPZCAT) and the corresponding $1\sigma$ error.

        \item SPEC\_Z: Spectroscopic redshift of the source, if available.

        \item T\_COUNTS/H\_COUNTS/S\_COUNTS: spectral background subtracted 
        counts in the full/hard/soft bands computed by adding all available 
        instruments and exposures for the corresponding observation.

        \item NHGAL: Galactic column density in the direction of the source 
        from the Leiden/Argentine/Bonn (LAB) Survey of Galactic HI.
    \end{itemize}

        \item \textbf{Model related columns} \\
        Columns referring to any particular model start with the model's name 
        (wapo, wamekal, wamekalpo, wapopo). 

        \begin{enumerate}
            \item Spectral-fit summary columns
            \begin{itemize}\setlength\itemsep{0.2em}
                \item A\_FIT: The value is set to True, if an acceptable fit, 
                i.e. KS p-value $> 0.01$, has been found for at least one of 
                the models applied, and to False otherwise.

                \item P\_MODEL: The data preferred model, that is the model with 
                the highest evidence (lowest logZ). A spectral model is always 
                listed regardless of the fit being an acceptable or an 
                unacceptable fit.

                \item A\_MODELS: List of acceptable models. This column 
                contains the remaining models with relative evidence (with 
                respect to P\_MODEL) lower than 30 ('very strong evidence' 
                accordingly to the scale of \citealt{Jeffreys61}). Assuming all 
                models \textit{a priori} equally probable, there is no statistical reason 
                to rule out any of the models in the set formed by P\_MODEL and 
                A\_MODELS. Hence, if the fit is acceptable, the simplest model 
                should be selected as the best-fit model.
            \end{itemize}

            \item \textit{Parameters and errors.} \\
            Columns referring to parameters and errors start with the model 
            name and the parameter name (logNH[1,2], PhoIndex[1,2], kT, and z). 
            Values for the normalizations of the models and the relative 
            normalization factors between instruments are not included in the 
            table (but they are available in the SQL database).

            \begin{itemize}\setlength\itemsep{0.2em}
                \item <MODEL>\_<PARAMETER>: parameter value.

                \item <MODEL>\_<PARAMETER>\_min, <MODEL>\_<PARAMETER>\_max: upper and    
                lower limits of the 90\% credible interval for the parameter.
            \end{itemize}

            \item \textit{Fluxes and luminosities.}

            \begin{itemize}\setlength\itemsep{0.2em}
                \item <MODEL>\_flux\_<BAND>: the mean observed flux (in erg 
                cm$^{-2}$ s$^{-1}$) of all instruments and exposures for the 
                corresponding observation, in <BAND> (soft/hard). Observed 
                fluxes were corrected of Galactic absorption.

                \item <MODEL>\_fluxmin\_<BAND>, <MODEL>\_fluxmax\_<BAND>: lower and 
                upper limits of the 90\% credible interval.

                \item <MODEL>\_intflux\_<BAND>: the mean intrinsic flux 
                (rest-frame, corrected of intrinsic absorption, in erg 
                cm$^{-2}$ s$^{-1}$) of all instruments and exposures for the 
                corresponding observation, in <BAND> (soft/hard).

                \item <MODEL>\_intfluxmin\_<BAND>, <MODEL>\_intfluxmax\_<BAND>: lower 
                and upper limits of the 90\% credible interval.

                \item <MODEL>\_lumin\_<BAND>: the mean luminosity (rest-frame, 
                corrected of intrinsic absorption, in erg s$^{-1}$) of all 
                instruments and exposures for the corresponding observation, in 
                <BAND> (soft/hard).

                \item <MODEL>\_luminmin\_<BAND>, <MODEL>\_luminmax\_<BAND>: lower and 
                upper limits of the 90\% credible interval.
            \end{itemize}
        
            \item \textit{Fitting statistics.}
        
            \begin{itemize}\setlength\itemsep{0.2em}
                \item <MODEL>\_wstat: W-stat (Cash statistics) value.

                \item <MODEL>\_dof: Degrees of freedom.

                \item <MODEL>\_ks: Kolmogorov-Smirnov (KS) statistic.

                \item <MODEL>\_ks\_pvalue: KS p-value.

                \item <MODEL>\_logZ: Natural logarithm of the evidence, estimated 
                by the MultiNest algorithm.
            \end{itemize}
	\end{enumerate}
\end{enumerate}
\end{appendix}

\end{document}